# Investigation on the temperature dependence of substrate-induced in-plane uniaxial magnetic anisotropy in Ni thin films grown on 128° Y-cut $LiNbO_3$

Kalani Perera[1,2], Morton Greenslit[2], Luke Doucette[2], Mauricio Pereira da Cunha[2,3], and Nicholas S. Bingham[1,2]

1. Department of Physics & Astronomy, University of Maine, Orono, ME USA, 04469
2. Frontier Institute for Research in Sensor Technologies (FIRST), University of Maine, Orono, ME USA, 04469
3. Department of Electric and Computer Engineering, University of Maine, Orono, ME USA, 04469

## Abstract

We report the temperature dependence of the substrate-induced magnetic anisotropy in continuous Ni thin films deposited at room temperature using magnetron sputtering on a 128° Y-cut $LiNbO_3$ substrate. Ultrathin films exhibit a pronounced temperature dependence in magnetization, with as-grown films showing isotropic behavior and an unconventional hysteresis branch crossing at relatively low temperatures, whereas no branch crossing is observed in annealed films over the investigated temperature range, indicating suppression of the underlying mechanism. Temperature-dependent magnetization measurements show that post-deposition annealing establishes uniaxial anisotropy with well-defined easy and hard axes, whose strength evolves with temperature due to the nature of residual strain governed by both lattice mismatch and coefficient of thermal expansion (CTE) mismatch between the Ni film and the 128° Y-cut substrate, independent of film thickness for both 5 nm and 100 nm films. The emergence of this anisotropy following annealing indicates enhanced magnetoelastic coupling at the film-substrate interface, as the magnetic response begins to follow the anisotropic strain imposed by lattice mismatch and CTE mismatch.

## I. Introduction

Magnetic anisotropy[1–3] and its temperature dependence are key parameters in the design and selection of materials for advanced technological applications. Modern efforts in magnetic sensing platforms,[1,2,4–9] magnetic tunnel junctions,[10,11] magnetic random access memories,[7,10,12] logic devices,[12,13] and magnonic devices,[2,14] aim to exploit controllable anisotropy to achieve efficient magnetic switching, miniaturization, and seamless integration to other functional architectures.[4,6,9,15]

Multiferroic materials are promising candidates for advanced functional devices because they exhibit two or more coexisting ferroic orders, e.g. ferroelectricity and ferromagnetism.[3,9,15–18] However, single-phase multiferroics are rare and exhibit weak coupling, typically below room temperature, resulting in limited tunability due to competing constraints between ferroic orders.[3,15,17–19] Recently, multiferroic[3,19,20] heterostructures have shown magnetoelectric coupling that is three or more orders[19] of magnitude stronger than single-phase multiferroics, enabled by strain-mediated coupling across the interface. Strategies to tune these responses include high-temperature growth during deposition, application of an *in situ* magnetic field during deposition, post-deposition annealing, film thickness, and material composition.[1,2,5,9,15,21] For optimal device performance, efficient strain transfer at the interface must be maximized.[19]

In thin films, the origin and strength of magnetic anisotropy depend on the interplay between several contributions, such as magnetocrystalline anisotropy,[4,21,22] shape anisotropy,[21] magnetoelastic anisotropy[4,5] from substrate-induced strain, magnetostriction,[4,5,10,15,16,21] residual stress,[4] and domain structure.[4]

Single crystal 128° Y-cut $LiNbO_3$ is a piezoelectric,[1,3,19,23] prototypical ferroelectric,[3,8,9,19,23] and pyroelectric,[8,23] that can enable tunable strain-mediated magnetoelectric and magnetoelastic coupling in ferromagnetic thin films.[1,2,4–6,10,19,21] Ni is a ferromagnetic 3d transition metal with a Curie temperature ($T_C$) of ~627 K,[1,10] which provides a robust platform for studying strain-controlled magnetic anisotropy in these heterostructures.

Although strain-mediated magnetic anisotropy in multiferroic heterostructures has been widely investigated, the temperature dependence of interfacial strain remains underexplored. Here, we investigate the temperature dependence of substrate-induced in-plane uniaxial magnetic anisotropy in Ni films grown on 128° Y-cut $LiNbO_3$ substrate, focusing on the influence of film-substrate coupling at the interface and the modifications induced by post-deposition annealing.

## II. EXPERIMENTAL METHODS

Ni films with nominal thicknesses of 5 nm and 100 nm were deposited on single-crystal 128° Y-cut $LiNbO_3$ substrate at room temperature using DC magnetron sputtering. The deposition was performed at Ar pressure of 3.0 mTorr and a DC power of 50W, with a base pressure of $\sim 5.0 \times 10^{-5}$ Torr. A 5 nm Ti capping layer was deposited superjacent to prevent deleterious oxidation states. All characterizations were performed in both the as-grown state and after annealing at 623 K for 30 minutes in vacuum, along two mutually perpendicular in-plane crystallographic directions of the substrate. For 128° Y-cut $LiNbO_3$ these in-plane directions correspond to $x = x' = (2\bar{1}.0)$ and $y' = (01.2)$.[1,24]

The crystal structures of the Ni films were characterized at 300K using grazing incidence X-ray diffraction (GIXRD, X'Pert Pro). Magnetic properties were measured over temperatures ranging from 10-200 K using superconducting quantum interference device magnetometer (SQUID, MPMS XL Quantum Design). Detailed temperature-dependent magnetic measurements were primarily performed on the 5 nm films, while the 100 nm films were measured to provide a comparison between ultrathin and thicker films. Attention was paid to the "saddling" of the sample to avoid improper alignment artifacts, particularly along the hard axis during resolved magnetization measurements.[25,26]

## III. RESULTS AND DISCUSSION

### A. Structural Properties (X-ray Diffraction)

The optimal incident angles (ω) for GIXRD scans for the as-grown 100 nm Ni film was ω = 0.9° and ω = 0.5° after annealing. This is attributed to structural evolution within film, including atomic arrangement, modifications in crystal ordering, and changes in the effective film density.[27–29] Whereas, variations in surface roughness, can modify the effective X-ray penetration depth in grazing incidence geometries.[29,30] These structural changes can depend on annealing conditions, including annealing temperature, environment, and duration.[27–30]

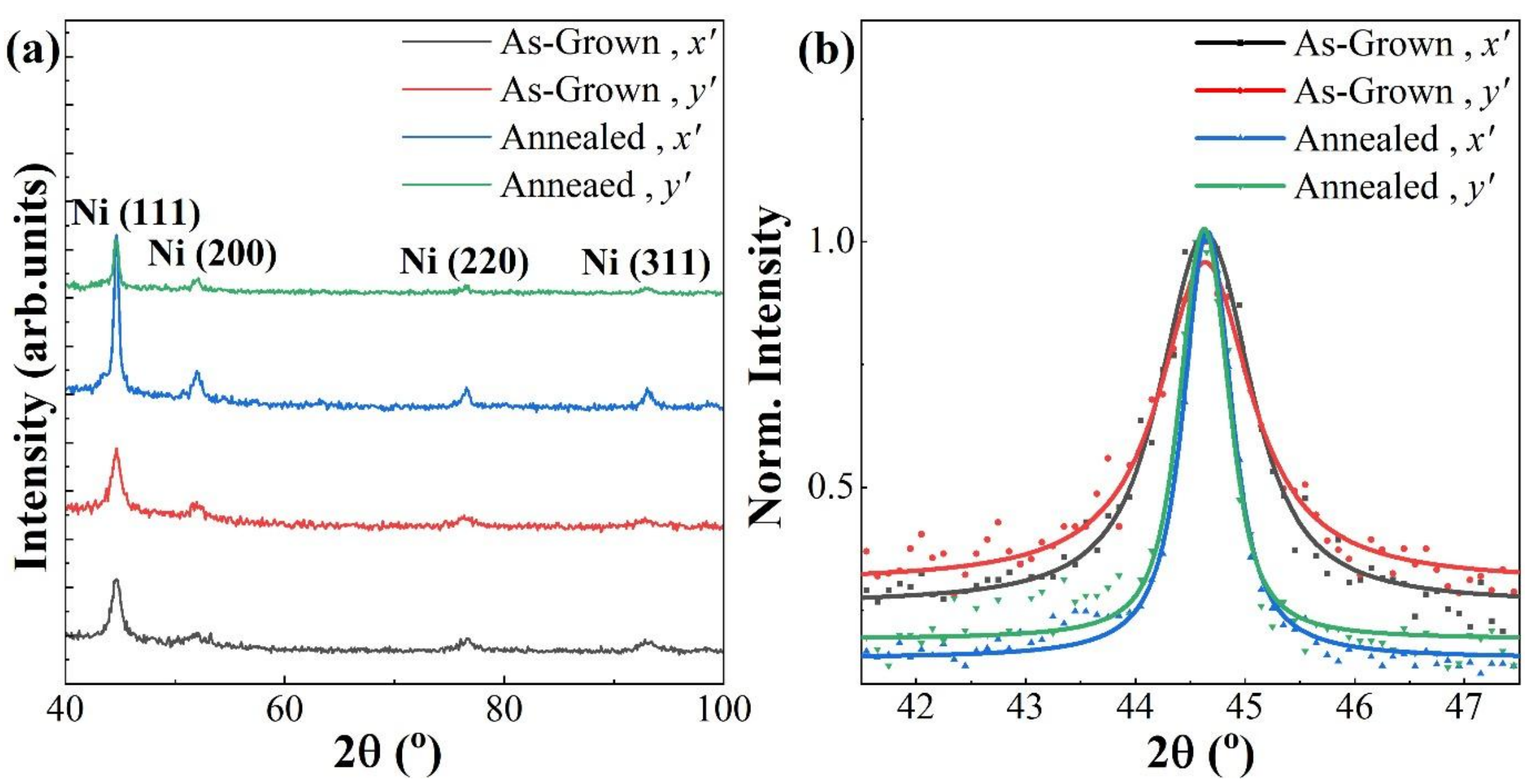

**FIG. 1 (a)** GIXRD for 100 nm Ni films deposited on 128° Y-cut $LiNbO_3$ and **(b)** zoomed in Ni (111) peak.

GIXRD measurements along both in-plane directions ($x'$ and $y'$) confirm a polycrystalline face-centered cubic (fcc) Ni structure in both as-grown and annealed films.[1] The Ni(111) peak exhibits the highest intensity, indicating that crystallites with the (111) orientation remain dominant after annealing. A decrease in the FWHM was observed for all measured peaks after annealing, indicating improved crystallinity and grain growth in the films.

The primary grain size was estimated using the Scherrer equation,[27,28,30]

$$d = \frac{k\lambda}{\beta cos\theta} \tag{1}$$

where $d$ is the grain size, $k \sim 0.9$ is the shape factor, λ is the X-ray wavelength, β the full-width-at-half maximum (FWHM) of the strongest diffraction peak, and θ is the Bragg diffraction angle of the corresponding peak.[27,28,30] For grains in the 10-20 nm range, the Scherrer equation works reasonably well, but absolute errors can be > 10 %.[27]

For the as-grown Ni films, the FWHM of the Ni (111) peak was approximately ~ 0.99° and ~ 0.98° along $x'$ and $y'$ directions, respectively, while for annealed films, they were ~ 0.49° along both $x'$ and $y'$ directions. The corresponding grain sizes increased from ~ 8 nm in the as-grown films to ~18 nm after annealing, indicating substantial grain growth and improved crystallinity induced by the post-deposition annealing process.

The approximate micro-lattice strain (ε) was estimated using the Williamson-Hall formula,[27]

$$\varepsilon = \frac{\beta}{4 \tan \theta} \tag{2}$$

The as-grown Ni films exhibited a lattice strain of approximately ε ~ 0.01, which decreased to ε~ 0.005 after post-deposition annealing. This reduction indicates that annealing promotes lattice relaxation in the Ni films. It should be noted that these values are approximate and should only be used for qualitative comparisons.

The absence of observable peak shifts indicates that the GIXRD peak positions remain unchanged within the penetration depth probed by the grazing incidence geometry, suggesting that any anisotropic strain evolution along orthogonal in-plane directions is below the detection sensitivity. Previously, Xiang et al.[4] reported that highly textured Ni films with thicknesses between 180 and 510 nm grown on PMN-PT (001) substrates exhibited a gradual decrease in in-plane tensile strain with increasing thickness, as revealed by reciprocal space mapping (RSM) and in-plane ferromagnetic resonance (FMR) measurements. The 180 nm films displayed the largest anisotropy field, indicating that the observed uniaxial magnetic anisotropy is likely a substrate-induced effect rather than intrinsic strain within the Ni layer.[4]

No distinct diffraction peaks were observed in the GIXRD measurements for 5 nm Ni films on 128° Y-cut $LiNbO_3$, which may be attributed to the combined effects of small crystallite size, structural disorder, limited scattering volume, and sensitivity limits of the measurement.

## B. Magnetic Properties (SQUID Magnetometry)

Applied magnetic field dependent magnetization (M-H) curves of 100 nm Ni films in the as-grown state and annealed state (annealed at 623 K for 30 minutes) were measured at 10 K and 200 K along the two orthogonal in-plane crystallographic directions ($x'$ and $y'$). In Fig. 2, the normalized magnetization for these measurements is shown to highlight the induced anisotropy and enable easy comparison between thick and thin films. M-H curves of the as-grown state were isotropic in both in-plane crystallographic directions. M-H curves of the post-deposition annealed films demonstrated uniaxial anisotropy with a well-defined easy axis along $x'$ and a hard axis along the $y'$ direction.

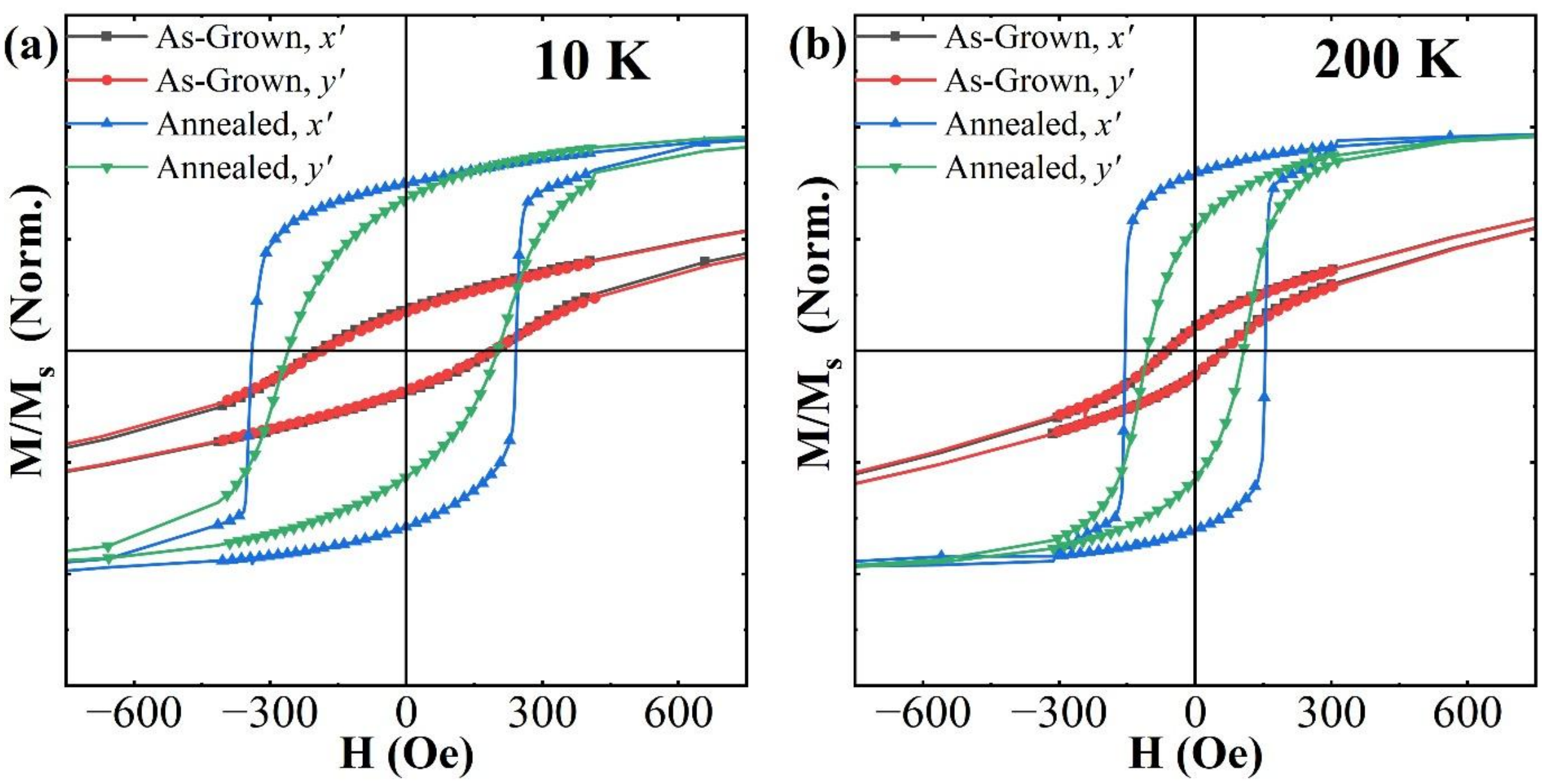

**FIG. 2.** M-H curves of 100 nm thick Ni films on 128° Y-cut $LiNbO_3$ at **(a)** 10 K, **(b)** 200 K.

The low remanence ratio (remanence ($M_r$)/ saturation magnetization ($M_s$)) and high saturation field observed in the as-grown state, exceeding the anisotropy field of the hard axis from annealing, are attributed to growth-induced residual strain and polycrystallinity. Residual strain in thin films primarily originates from CTE mismatch, lattice mismatch, and interface misfit dislocations, where misfit dislocations partially relieve mismatch-induced strain.[31] Previous theoretical models and experimental studies have shown that for large lattice mismatch strain (>1%), growth transitions from layer-by-layer growth to island growth, leading to substantially lower critical film thickness (< ~10 nm) for the nucleation of misfit dislocations.[31]

GIXRD results of the 100 nm Ni films showed fcc Ni growth with strong Ni (111) orientation, indicating that Ni(111) planes are most likely preferentially oriented parallel to the substrate surface. For fcc Ni with lattice constant $a_{Ni}$ =3.52 Å[32] at room temperature, while the lattice parameter in the Ni (111) plane is 2.49Å. The hexagonal basal planes of $LiNbO_3$ have lattice parameters a = 5.1489 Å, c =13.8631 Å at room temperature.[33] Due to the 128° Y-cut geometry, the two orthogonal in-plane directions correspond to different projections of the hexagonal lattice, resulting in anisotropic in-plane periodicities of $a_X$ ≈5.1489 Å and $a_Y$≈10.56 Å, estimated from the Euler-rotated crystal structure. This results in significant anisotropic lattice mismatch along two in-plane directions, which also varies with the temperature. Ni exhibits a CTE of ~13 ppm/K,[5,34,35] whereas 128° Y-cut $LiNbO_3$ shows anisotropic CTE values with $\alpha_{x'} \approx 14.8$ ppm/K and $\alpha_{y'} \approx 8.2$ ppm/K at 300 K.[5,14] According to literature reports on the CTE behavior of $LiNbO_3$, the CTE mismatch along the $x'$ direction remains smaller than along the $y'$ direction.[14] Between 300 K and 10 K, the CTE of Ni decreases steadily without pronounced fluctuations, which exhibit a generally monotonic, quasi-linear behavior, becoming small relative to room temperature values and approaching zero with temperature.[14,34,35] At approximately 60 K, $LiNbO_3$ undergoes a phase transition that modifies the CTE trend along $y'$ crystallographic direction.[14] Aside from this, thermal strain at low temperatures is small. Nevertheless, the strain accumulated during growth and cooling persists, contributing to the residual strain observed in films.[14]

Although these anisotropic strain contributions are expected to induce uniaxial magnetic anisotropy along $x'$ and $y'$ crystallographic directions, the as-grown films exhibit isotropic behavior in M-H curves at both 10 K [Fig. 2 (a)] and 200 K [Fig. 2 (b)]. This behavior can be attributed to kinetically limited growth conditions during the magnetron sputtering. Atoms arriving at the substrate are incorporated with limited surface diffusion, preventing relaxation into energetically favorable lattice configurations,[36] where the kinetic constraints on adatom mobility are strongly governed by deposition conditions. In addition, substrate heating up to approximately 200 K arises from energetic atom bombardment and heat released during the condensation.[36,37] These conditions lead to small crystallites, high defect density, and randomly oriented grains.[36] Although lattice and CTE mismatch are expected to induce uniaxial anisotropy in principle, the as-grown films exhibit isotropic M-H behavior along two in-plane axes of the $LiNbO_3$ substrate, suggesting that as-grown films exhibit weak magnetoelastic coupling. Furthermore, with increasing thickness, strain relaxation via misfit dislocations further disrupts strain uniformity, reinforcing the absence of a well-defined anisotropy in the as-grown state.[31]

Upon annealing at 623 K for 30 minutes, a well-defined uniaxial magnetic anisotropy is established, with distinct easy and hard axes. The easy axis is observed along the $x'$ crystallographic direction, where the CTE mismatch between the film and substrate is relatively small, while the hard axis aligns along the $y'$ direction, where the mismatch is significantly larger. This anisotropy is consistent with the emergence of substrate-induced anisotropy, where the reduced mismatch along the less strained $x'$ favors the preferential crystallite alignment, and consequently, the larger mismatch along $y'$ gives rise to the hard axis. Compared to the as-grown films, the hard axes exhibit higher coercivity ($H_c$), $M_r$, and low saturation magnetic field, indicating both the relaxation of residual strain and the improved crystalline quality, and in-plane magnetic moment realignment as confirmed by M-H curves and GIXRD scans. Higher $H_c$ observed at 10 K [Fig. 2(a)] relative to 200 K [Fig. 2(b)] reflects that the thermal energy at elevated temperatures facilitates magnetization reversal. In addition to strain relaxation through post-deposition annealing, the CTE mismatch and lattice mismatch appear to promote anisotropic reorientation of crystallites during annealing, resulting in substrate-induced magnetic anisotropy through magnetoelastic coupling. The persistence of this anisotropy at both 10 K and 200 K suggests that substrate-induced magnetoelastic effects constitute a dominant contribution to the magnetic behavior of the relatively thick (100 nm) ferromagnetic films.

Ultrathin films exhibit finite-size effects and strong interfacial influence, causing their ferromagnetic response to differ significantly from that of bulk or thicker films. Factors such as lattice defects, quantum size effects, interface-driven electronic reconstruction, intermixing, and the formation of magnetically “dead” layers can modify magnetic anisotropy, highlighting the dominant role of the substrate in determining the magnetic behavior at this scale.[7,15,21]

From Fig 3 (a), the M-H curves measured at 10 K for the as-grown 5 nm Ni films exhibit nearly isotropic behavior. Compared to the post-deposition annealed state, the as-grown state displays higher $H_c$ and $M_r/M_s$ than the annealed hard axis. $M_r/M_s$ of the as-grown state approaches that of the easy axis, while the saturation field remains lower than that of the annealed hard axis. This suggests the Ni films are experiencing stronger substrate-induced effects near the interface. This behavior can be understood in terms of the combined effects of large lattice mismatch, large CTE mismatch, and possible absence of misfit dislocations in an ultrathin film.[31] The lattice mismatch (>1%) promotes island growth, leading to small crystallite sizes, structural disorder, and random moment orientation. Additionally, due to the nature of magnetron sputtering, atoms do not undergo sufficient surface diffusion to reach an energetically favorable lattice configuration that accommodates the film-substrate mismatch in as-grown films.[36,37] Therefore, as-grown films experience strong growth-induced residual strain, which appears not to be strongly mediated through magnetoelastic coupling, even at the interface, similar to the behavior observed in 100 nm films.

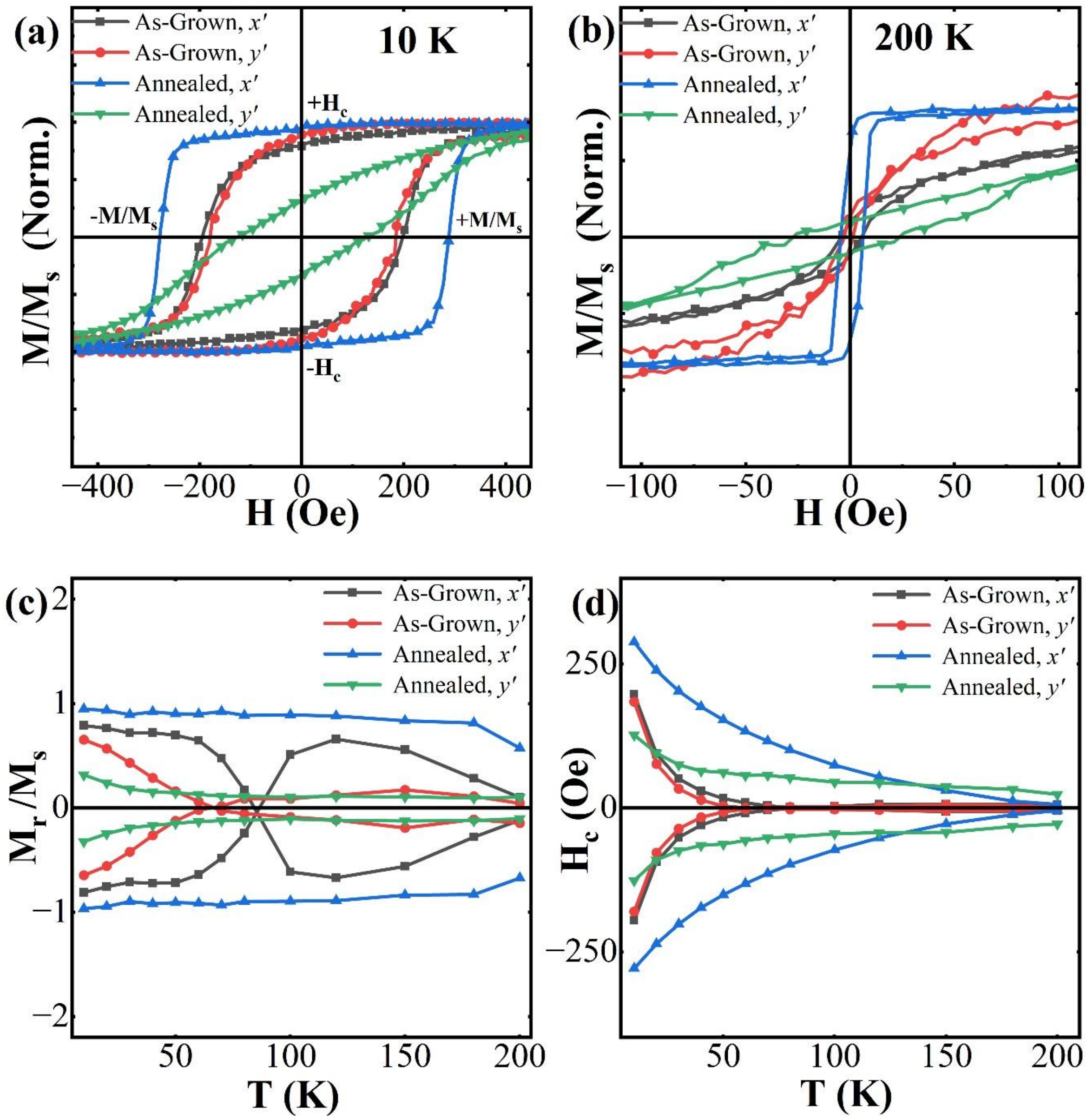

**FIG. 3.** M-H curves of 5 nm Ni films on 128° Y-cut $LiNbO_3$ at **(a)** 10 K, **(b)** 200 K **(c)**Temperature dependence of $M_r/M_s$ over the range 10-200 K for 5 nm Ni films and **(d)** $H_c$ over the range 10-200 K for 5 nm Ni films.

Fig. 3 (c) and Fig. 3 (d), derived from M-H curves from 10K to 200 K, reveal the temperature-dependent evolution of magnetic behavior. Along $x'$ crystallographic direction, $M_r/M_s$ remains near ~1.0 to ~ 60 K before branch crossing near 86 K. Along $y'$ crystallographic direction, $M_r/M_s$ decays gradually from 10 K, with branch crossing occurring ~60 K, this phenomenon is attributed to the high anisotropy in the film.[2,13] Despite these differences in $M_r$ evolution, $H_c$ remains nearly identical along both directions and decreases with increasing temperature. This can be attributed to the lattice and CTE mismatch which are smaller along the $x'$ crystallographic direction, providing partial stabilization of magnetic moments at low temperatures, whereas the larger mismatch along $y'$ crystallographic direction leads to early, gradual misalignment due to higher residual strain. However, the weak and incoherent magnetoelastic coupling, combined with small, randomly oriented crystallites, results in similar energy barriers for magnetization reversal along both axes, producing nearly isotropic $H_c$. These observations indicate that, in the as-grown state, despite the film thickness, substrate-induced anisotropic effects are locally present but insufficient to overcome growth-induced residual strain and generate a well-defined global uniaxial anisotropy, which evolves more rapidly in ultrathin films.

Once branch crossing occurs, the magnetization reversal appears to deviate from conventional ferromagnetic behavior, in which the moments no longer follow the applied field; instead, they appear to rotate away from the field at low applied fields. This behavior likely arises from the combined effects of high local anisotropy induced by growth-related residual strain, which remains unrelieved due to the absence of misfit dislocations, unlike thicker films, and increased thermal energy at elevated temperatures, which allows magnetic moments to overcome weak pinning and partially reorient before the applied field is sufficient to align them. Mathews *et al.*[13] modified the Stoner–Wohlfarth (S-W) model to describe this branch crossing, incorporating temperature dependence and grain-level anisotropy variation absent in the idealized model.[13,25]

After annealing, a well-defined uniaxial magnetic anisotropy is established, with distinct easy($x'$) and hard($y'$) axes. Even at 200 K [Fig. 3 (b)], the M-H curves do not exhibit branch crossing; however, the $H_c$ is significantly diminished, suggesting that further increase in temperature could eventually induce branch crossing. This reduction in $H_c$ is more pronounced in the 100 nm films, indicating a stronger temperature sensitivity in the ultrathin regime. From Fig. 3 (d), the $H_c$ along the easy axis decreases rapidly with increasing temperature, while along the hard axis remains relatively higher, reflecting the persistence of anisotropic energy barriers. In contrast, $M_r/M_s$ along the easy axis remains close ~1.0, whereas along the hard axis it decreases and approaches conditions associated with branch crossing at 200 K.

These trends indicate that the magnetization behavior in thin films is strongly temperature dependent, even after the preferential alignment of magnetic moments along energetically favorable directions defined by lattice and CTE mismatch between the film substrate, as well as crystallite growth, reorientation, and modifications in defect density. The absence of hysteresis branch crossing in the M-H curves is attributed to the reduction of residual strain upon annealing, which promotes morphological ordering and the emergence of uniaxial anisotropy, indicating effective film-substrate coupling through interfacial magnetoelastic interactions.

As effects such as residual strain, the nature of crystallinity of the ferromagnetic thin layer, and the temperature at which the application operates determine the nature of the magnetoelastic coupling effect, ultrathin Ni films may become less suitable for certain applications despite offering higher tunability compared to bulk counterparts.[4,10]

## IV. CONCLUSION

Ni films grown on 128° Y-cut $LiNbO_3$ via room temperature magnetron sputtering exhibit almost isotropic magnetic behavior in the as-grown state, despite anisotropic lattice and CTE mismatch along two orthogonal in-plane crystallographic directions. This indicates that the rapid and kinetically limited nature of magnetron sputtering leads to significant growth-induced residual strain, polycrystallinity, and weak film-substrate coupling, thereby suppressing effective magnetoelastic coupling even at ultrathin thicknesses.

Post-deposition annealing promotes structural relaxation, crystallite growth, and magnetic moment reorientation, leading to a well-defined uniaxial anisotropy. This is attributed to strain relaxation and strain redistribution driven by lattice and CTE mismatch, which enhances magnetoelastic coupling at the film-substrate interface, and is observed in both ultrathin and thicker films.

However, the temperature evolution of the magnetic properties differs significantly with thickness. Ultrathin films exhibit strong temperature dependence and unconventional hysteresis branch crossing, indicating that thermal energy contribution, high growth-induced residual strain, and possibly competing local anisotropies are dominating the magnetic response. In contrast, thicker (100 nm) films show more stable magnetic behavior with low temperature sensitivity, reflecting more coherent and robust substrate-induced magnetoelastic coupling.

## ACKNOWLEDGMENTS

This material is based upon work supported by the U.S. Department of Energy, Office of Science, Office of Basic Energy Sciences, Established Program to Stimulate Competitive Research (EPSCoR) under Award Number DE-SC0021981

## REFERENCES

[1] M. Ito, S. Ono, H. Fukui, K. Kogirima, N. Maki, T. Hikage, T. Kato, T. Ohkochi, A. Yamaguchi, M. Shima, and K. Yamada, "Uniaxial in-plane magnetic anisotropy mechanism in Ni, Fe, and Ni-Fe alloy films deposited on single crystal Y-cut 128° LiNbO3 using magnetron sputtering," Journal of Magnetism and Magnetic Materials **564**, 170177 (2022).

[2] S.A. Mathews, and J. Prestigiacomo, "Controlling magnetic anisotropy in nickel films on LiNbO3," Journal of Magnetism and Magnetic Materials **566**, 170314 (2023).

[3] M. Staruch, D.B. Gopman, Y.L. Iunin, R.D. Shull, S.F. Cheng, K. Bussmann, and P. Finkel, "Reversible strain control of magnetic anisotropy in magnetoelectric heterostructures at room temperature," Sci Rep **6**(1), 37429 (2016).

[4] Y. Xiang, K. Liang, S. Keller, M. Guevara, M. Sheng, Z. Yan, P. Zhou, Y. Qi, Z. Ma, Y. Liu, G. Srinivasan, G.P. Carman, T. Zhang, and C.S. Lynch, "Thickness-dependence of magnetic anisotropy and domain structure in Ni thin films grown on a PMN-PT substrate," Smart Mater. Struct. **29**(9), 095019 (2020).

[5] S.A. Mathews, N.S. Bingham, R.J. Suess, K.M. Charipar, R.C.Y. Auyeung, H. Kim, and N.A. Charipar, "Thermally Induced Magnetic Anisotropy in Nickel Films on Surface Acoustic Wave Devices," IEEE Trans. Magn. **55**(2), 1–4 (2019).

[6] M. Gruber, R. Konetschnik, M. Popov, J. Spitaler, P. Supancic, D. Kiener, and R. Bermejo, “Atomistic origins of the differences in anisotropic fracture behaviour of LiTaO3 and LiNbO3 single crystals,” Acta Materialia **150**, 373–380 (2018).

[7] Tao Wu, A. Bur, J.L. Hockel, Kin Wong, Tien-Kan Chung, and G.P. Carman, “Electrical and Mechanical Manipulation of Ferromagnetic Properties in Polycrystalline Nickel Thin Film,” IEEE Magn. Lett. **2**, 6000104–6000104 (2011).

[8] M.V. Jacob, J.G. Hartnett, J. Mazierska, V. Giordano, J. Krupka, and M.E. Tobar, “Temperature Dependence of Permittivity and Loss Tangent of Lithium Tantalate at Microwave Frequencies,” IEEE Trans. Microwave Theory Techn. **52**(2), 536–541 (2004).

[9] R. Gao, J. Xue, H. Wu, Y. Ye, J. Wang, and Q. Liu, “Acoustically driven spin wave resonance in Ni/128° Y-cut LiNbO3 hybrid devices with the beam steering effect,” Applied Physics Letters **123**(23), 232401 (2023).

[10] S. Finizio, M. Foerster, M. Buzzi, B. Krüger, M. Jourdan, C.A.F. Vaz, J. Hockel, T. Miyawaki, A. Tkach, S. Valencia, F. Kronast, G.P. Carman, F. Nolting, and M. Kläui, “Magnetic Anisotropy Engineering in Thin Film Ni Nanostructures by Magnetoelastic Coupling,” Phys. Rev. Applied **1**(2), 021001 (2014).

[11] A.S. Goossens, K. Samanta, A. Jaman, W. Boubaker, J.J.L. Van Rijn, E.Y. Tsymbal, and T. Banerjee, “Symmetry-driven large tunneling magnetoresistance

in SrRuO 3 magnetic tunnel junctions with perpendicular magnetic anisotropy," Phys. Rev. Materials **8**(9), L091401 (2024).

[12] J.-M. Hu, C.-W. Nan, and L.-Q. Chen, "Size-dependent electric voltage controlled magnetic anisotropy in multiferroic heterostructures: Interface-charge and strain comediated magnetoelectric coupling," Phys. Rev. B **83**(13), 134408 (2011).

[13] S.A. Mathews, A.C. Ehrlich, and N.A. Charipar, "Hysteresis branch crossing and the Stoner–Wohlfarth model," Sci Rep **10**(1), 15141 (2020).

[14] J.S. Browder, and S.S. Ballard, "Thermal expansion data for eight optical materials from 60 K to 300 K," Appl. Opt. **16**(12), 3214 (1977).

[15] F. Maccherozzi, M. Ghidini, M. Vickers, X. Moya, S.A. Cavill, H. Elnaggar, A.D. Lamirand, N.D. Mathur, and S.S. Dhesi, "Inverted shear-strain magnetoelastic coupling at the Fe/BaTiO3 interface from polarised x-ray imaging," Nat Commun **16**(1), 8445 (2025).

[16] A. Arora, L.C. Phillips, P. Nukala, M. Ben Hassine, A.A. Ünal, B. Dkhil, Ll. Balcells, O. Iglesias, A. Barthélémy, F. Kronast, M. Bibes, and S. Valencia, "Switching on superferromagnetism," Phys. Rev. Materials **3**(2), 024403 (2019).

[17] D.K. Pradhan, S. Kumari, R.K. Vasudevan, E. Strelcov, V.S. Puli, D.K. Pradhan, A. Kumar, J.M. Gregg, A.K. Pradhan, S.V. Kalinin, and R.S. Katiyar, "Exploring the Magnetoelectric Coupling at the Composite Interfaces of FE/FM/FE Heterostructures," Sci Rep **8**(1), 17381 (2018).

[18] A. Ghaffar, M. Fatima, G.M. Mustafa, S.A. Hafeez, A. Mahmood, and S. Atiq, "Tunability of magnetoelectric coupling in perovskite/spinel-based triphasic multiferroic composites for multistate devices," Ceramics International **49**(7), 10298–10304 (2023).

[19] A. Begué, and M. Ciria, "Strain-Mediated Giant Magnetoelectric Coupling in a Crystalline Multiferroic Heterostructure," ACS Appl. Mater. Interfaces **13**(5), 6778–6784 (2021).

[20] J. Shin, S. Hoon Kim, Y. Suwa, S. Hashi, and K. Ishiyama, "Control of in-plane uniaxial anisotropy of Fe72Si14B14 magnetostrictive thin film using a thermal expansion coefficient," Journal of Applied Physics **111**(7), 07E511 (2012).

[21] S. Wang, L. Hu, G.D. Zhang, R.H. Wei, W.H. Song, X.B. Zhu, and Y.P. Sun, "Continuous strain-mediated perpendicular magnetic anisotropy in epitaxial (111) CoFe2O4 thin films grown on LiTaO3 substrates," Journal of Applied Physics **135**(12), 123905 (2024).

[22] A.A. El-Gendy, M. Qian, Z.J. Huba, S.N. Khanna, and E.E. Carpenter, "Enhanced magnetic anisotropy in cobalt-carbide nanoparticles," Appl. Phys. Lett. **104**(2), 023111 (2014).

[23] T. Abe, S. Shikano, K. Shimamura, H. Sugiyama, S. Ono, M. Shima, and K. Yamada, "Effect of crystal structure on in-plane magnetic anisotropy in cobalt thin films deposited on $LiTaO_3$ substrates," Jpn. J. Appl. Phys. **64**(11), 11SP35 (2025).

[24] S. Sanna, and W.G. Schmidt, "Lithium niobate X -cut, Y -cut, and Z -cut surfaces from *ab initio* theory," Phys. Rev. B **81**(21), 214116 (2010).

[25] S.A. Mathews, J. Prestigiacomo, and P.C. Mulford, "Hysteresis branch crossing in permalloy," Journal of Magnetism and Magnetic Materials **588**, 171454 (2023).

[26] Jin Hanmin, S. Dongsheng, G. Cunxu, and H. Kim, "Inverted hysteresis loops: Experimental artifacts arising from inappropriate or asymmetric sample positioning and the misinterpretation of experimental data," Journal of Magnetism and Magnetic Materials **308**(1), 56–60 (2007).

[27] S. Jana, S.C. Mishra, A. Debnath, P. Veerender, A. Das, V. Jayakrishnan, A. Das, J. Kishor, A.K. Chauhan, and D. Bhattacharya, "Influence of the annealing conditions on the structural, morphological and nanomechanical properties of co-sputter deposited NiTi films," Thin Solid Films **826**, 140755 (2025).

[28] B. Sun, P. Wang, B. Ma, J. Deng, and J. Luo, "Effects of annealing on the temperature coefficient of resistance of nickel film deposited on polyimide substrate," Vacuum **160**, 18–24 (2019).

[29] C. Wiemer, S. Ferrari, M. Fanciulli, G. Pavia, and L. Lutterotti, "Combining grazing incidence X-ray diffraction and X-ray reflectivity for the evaluation of the structural evolution of HfO2 thin films with annealing," Thin Solid Films **450**(1), 134–137 (2004).

[30] W.-E. Fu, Y.-Q. Chang, Y.-C. Chen, E.M. Secula, D.G. Seiler, R.P. Khosla, D. Herr, C. Michael Garner, R. McDonald, and A.C. Diebold, "Post-Deposition

Annealing Analysis for HfO[sub 2] Thin Films Using GIXRR/GIXRD," in *AIP Conference Proceedings*, (AIP, Albany (New York), 2009), pp. 122–126.

[31] A. Moridi, H. Ruan, L.C. Zhang, and M. Liu, "Residual stresses in thin film systems: Effects of lattice mismatch, thermal mismatch and interface dislocations," International Journal of Solids and Structures **50**(22–23), 3562–3569 (2013).

[32] W. Tian, H.P. Sun, X.Q. Pan, J.H. Yu, M. Yeadon, C.B. Boothroyd, Y.P. Feng, R.A. Lukaszew, and R. Clarke, "Hexagonal close-packed Ni nanostructures grown on the (001) surface of MgO," Applied Physics Letters **86**(13), 131915 (2005).

[33] Y.S. Kim, and R.T. Smith, "Thermal Expansion of Lithium Tantalate and Lithium Niobate Single Crystals," Journal of Applied Physics **40**(11), 4637–4641 (1969).

[34] F.C. Nix, and D. MacNair, "The Thermal Expansion of Pure Metals: Copper, Gold, Aluminum, Nickel, and Iron," Phys. Rev. **60**(8), 597–605 (1941).

[35] R.J. Corruccini, and J.J. Gniewek, *Thermal Expansion of Technical Solids at Low Temperatures; a Compilation from the Literature*, 0 ed. (National Bureau of Standards, Gaithersburg, MD, 1961), p. NBS MONO 29.

[36] J.T. Gudmundsson, "Physics and technology of magnetron sputtering discharges," Plasma Sources Sci. Technol. **29**(11), 113001 (2020).

[37] V.I. Shapovalov, A.E. Komlev, A.S. Bondarenko, P.B. Baykov, and V.V. Karzin, "Substrate heating and cooling during magnetron sputtering of copper target," Physics Letters A **380**(7–8), 882–885 (2016).